%% file: main.tex
\pdfoutput=1

\documentclass{article}
\usepackage[utf8]{inputenc}   

\usepackage{graphicx} 
\usepackage{epstopdf}
\usepackage{float} 
\usepackage{subcaption}
\usepackage[export]{adjustbox} 
\usepackage[english]{babel}
\usepackage{amsmath}
\usepackage{authblk}
\usepackage[utf8]{inputenc} 

\usepackage[letterpaper,top=2cm,bottom=2cm,left=3cm,right=3cm,marginparwidth=1.75cm]{geometry}

\usepackage[utf8]{inputenc}
\usepackage[T1]{fontenc}
\DeclareUnicodeCharacter{2212}{-}   
\DeclareUnicodeCharacter{2013}{--}  
\DeclareUnicodeCharacter{2014}{---} 
\DeclareUnicodeCharacter{03C0}{\ensuremath{\pi}} 

\usepackage{xr-hyper}
\usepackage{hyperref}
\newcommand{\rubpy}{[Ru(bpy)$_3$]$^{2+}$}
\newcommand{\rubpz}{[Ru(bpy)$_2$(bpz)]$^{2+}$}

\title{Tracking Electron, Proton, and Solvent Motion in Proton-Coupled Electron Transfer with Ultrafast X-rays}

\input{authors.tex}

\date{}
\begin{document}
\maketitle

\begin{abstract}

Proton-coupled electron transfer (PCET) is foundational to catalysis, bioenergetics, and energy conversion, yet capturing and disentangling the coupled motions of electrons, protons, and solvent has remained a major experimental challenge. We combine femtosecond optical spectroscopy, site-specific ultrafast soft X-ray absorption spectroscopy, and time-resolved X-ray scattering with advanced calculations to disentangle the elementary steps of PCET in solution. Using a ruthenium polypyridyl model complex, we directly resolve photoinduced electron redistribution, ligand-site protonation within $\sim$100 ps, and the accompanying solvent reorganization. This unified multimodal approach provides an orbital-level, atomistic picture of PCET, showing how electronic, nuclear, and solvation degrees of freedom can be separated experimentally. Our results establish a general X-ray framework for understanding and ultimately controlling PCET in catalysis, artificial photosynthesis, and biological energy flow.

\end{abstract}

\section{Introduction}

Proton-coupled electron transfer (PCET) lies at the heart of chemical energy conversion, enabling catalytic redox processes, enzyme function, nutrient cycling, and biofuel production by coupling charge transport to proton motion ~\cite{Tyburski, Weinberg, Nocera}.  In solution, particularly in aqueous and biological environments, solvation and hydrogen-bonding networks strongly modulate these reactions, for instance by stabilizing transient intermediates or mediating proton transfer through dynamic relays. Despite decades of study, directly observing how electrons, protons, and solvent molecules move in concert has remained a fundamental challenge~\cite{Hsieh,Tyburski}. Traditional methods, such as thermodynamic analyses ~\cite{Mayer,Agarwal2022}, kinetic isotope effects ~\cite{Zhang}, and rate-law studies ~\cite{Cui}, offer only indirect evidence and cannot unambiguously characterize the underlying PCET pathways. Optical and vibrational spectroscopies can indirectly infer proton transfer from vibrational or electronic spectral changes, but these signals are often convoluted with bulk solvent contributions and lack site specificity. 

Time-resolved X-ray methods have emerged as powerful tools for overcoming these limitations. Core-level spectroscopies provide element specificity and site selectivity, and recent computational and experimental studies have demonstrated the utility of ultrafast soft X-ray absorption spectroscopy (XAS) for capturing proton transfer dynamics in liquids and small molecules~\cite{Young2020, loe2021, soley2022, Nimmrich2024, C9CP05677G, D0SC00840K, Yin2023,ge2024, BenVan, Eckert, Norrel}. Time-resolved X-ray solution scattering (XSS) has been successfully applied to monitor  photoinduced intramolecular structural rearrangements and the coupled solvent reorganization~\cite{Kim2015,Biasin2016,vanDriel2016}, including  solute-solvent H-bonding dynamics~\cite{BiasinNatChem}, in real time. However, no prior work has simultaneously provided a coherent, atomistic picture of electron, proton, and solvent motions during PCET in solution. Such a picture is crucial for resolving fundamental mechanistic questions, such as sequential versus concerted PCET, and synchronous versus asynchronous motion~\cite{Goings2020, Tyburski, Agarwal2022,Morris2017,Nocera2}, and for providing design criteria for solvent environments that optimize reactivity~\cite{Drummer2022}.

We combine femtosecond optical spectroscopy, N K-edge XAS, and XSS, together with advanced simulations to investigate PCET in the model complex \rubpz{}, where bpy is 2,2$^\prime$-bipyridine and bpz is 2,2$^\prime$-bipyrazine (Figure~\ref{fig:Figure 1}A).  This multimodal approach allows us to disentangle the timescales and electronic changes due to charge transfer and protonation at a ligand site, and capture the solvent reorganization coupled to ligand protonation.  Ultimately, our work establishes a comprehensive, atomistic framework for understanding light-driven PCET in complex molecular systems.

\section{PCET Dynamics}
\rubpz{} belongs to a class of transition metal complexes widely used to study light-driven PCET~\cite{Wenger, Wenger2013, Gagliardi}, and its optical properties are well-established~\cite{Rilemma, Shinozaki, Sun}. Its electronic absorption spectrum shows two main MLCT bands: a higher energy transition at 413 nm, where the photoexcited electron localizes on the bpy ligands (MLCT$_{\text{bpy}}$), and a lower energy transition at 474 nm, where the electron localizes on the bpz ligand (MLCT$_{\text{bpz}}$). Emission occurs exclusively from the lower energy $^{3}$MLCT${_\text{bpz}}$ excited-state. The $^{3}$MLCT$_{\text{bpz}}$ emission is quenched in the presence of acid due to the protonation of the excited-state, with a reported pKa$^{*}$ of 3.5~\cite{Rilemma}, whereas the ground-state protonation occurs with a reported pKa of -0.15~\cite{Shinozaki}. Based on these pKa values, pH 1 is ideal for studying photoinduced PCET in \rubpz (Figure S1), offering minimal ground-state protonation, but efficient excited-state protonation, as confirmed by  steady-state optical measurements (Figure S2).

\begin{figure}[H]
\centering
\includegraphics[width=\linewidth]{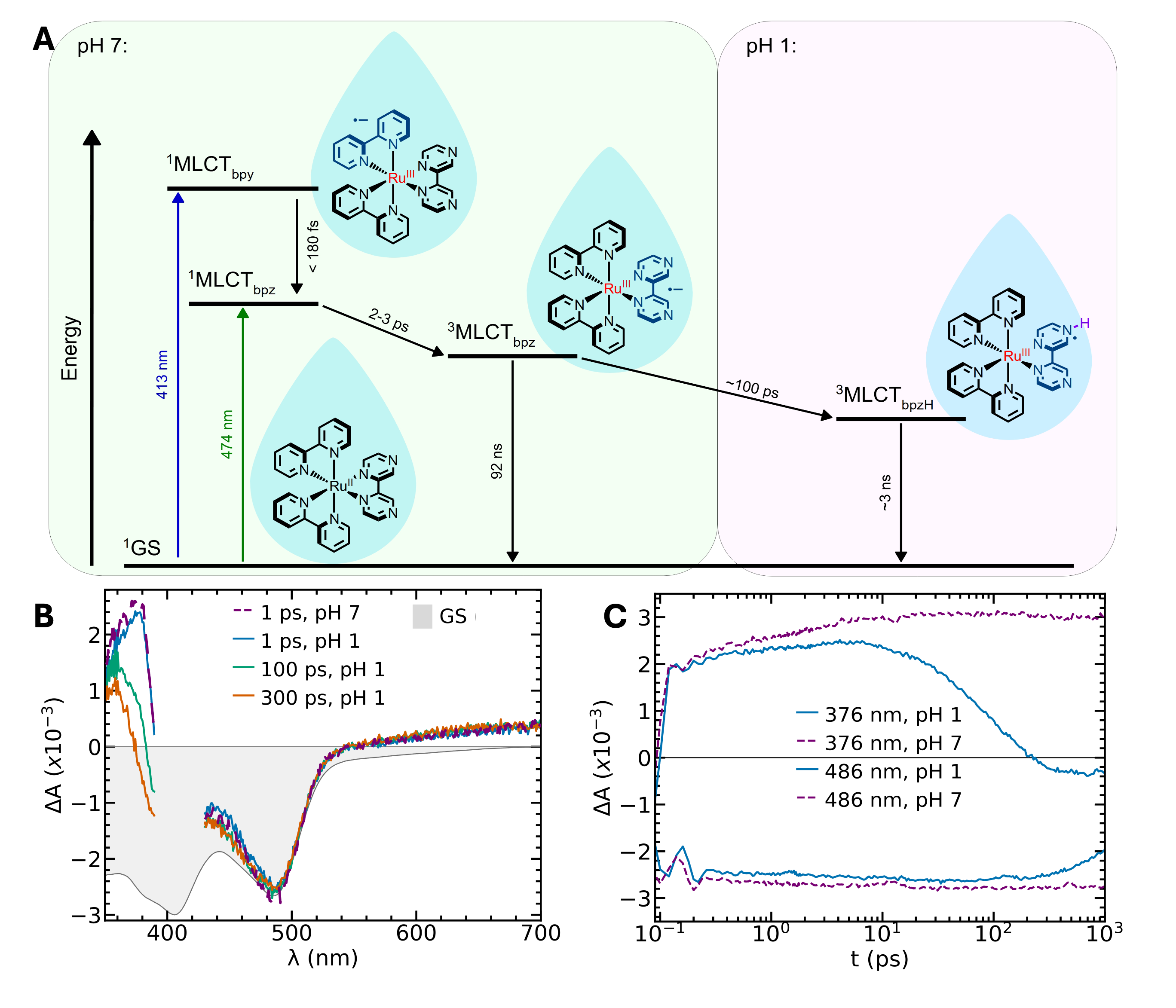}
\caption{(A) Overview of the excited-state dynamics of \rubpz{} in de-ionized water (pH 7) and 0.1 M HCl solution (pH 1). In acidic conditions, we probe the timescale of protonation with optical transient absorption (OTA), and capture the site-specific electronic dynamics and the coupled solvent reorganization with X-ray spectroscopies.  (B) OTA spectra measured 1 ps after 400 nm excitation of \rubpz{} in neutral and acidic conditions. OTA spectra measured 100 and 300 ps after photoexcitation at pH 1 showing spectral evolution due to excite state protonation of the $^{3}$MLCT${_\text{bpz}}$ state. The gray-shaded area shows the inverted ground-state UV-Vis absorption spectrum. (C) Kinetic traces measured at the excited-state absorption (376 nm) and ground-state bleach (486) show different excited-state dynamics at pH 1 and pH 7.}
\label{fig:Figure 1}
\end{figure}

Optical transient-absorption (OTA) measurements in de-ionized water (hereafter pH 7)  and 0.1 M HCl aqueous solution (hereafter pH 1) were conducted with an instrument response function (IRF) of $\sim$180 fs. Excitation at 400 nm (into MLCT$_{\text{bpy}}$) and 500 nm (into MLCT$_{\text{bpz}}$) produces identical OTA spectra and kinetics, indicating rapid inter-ligand electronic redistribution (within the IRF) to populate the long-lived $^{3}$MLCT${_\text{bpz}}$ state under both neutral and acidic conditions (Figure S3). Figure~\ref{fig:Figure 1}B shows the OTA spectrum of the $^{3}$MLCT$_{\text{bpz}}$ excited-state measured 1 ps after excitation. The spectrum exhibits ground-state bleach (GSB) of both MLCT bands, an excited-state absorption (ESA) centered at $\sim$350 nm, and a weak positive transient extending into the near-infrared region. 

At pH 7, the $^{3}$MLCT$_{\text{bpz}}$ state persists beyond the experimental time window ($\sim$1.4 ns), consistent with the reported 92 ns emission lifetime of \rubpz{} in water~\cite{Sun}. Kinetic traces (Figure~\ref{fig:Figure 1}C) show some minor spectral reshaping consistent with vibrational energy relaxation (VR) at early timescales, but no other excited-state dynamics.  

At pH 1, the early dynamics mirror those at pH 7, followed by spectral evolution over the first few hundred picoseconds (Figure ~\ref{fig:Figure 1}C). Global analysis~\cite{GlobalA} of the pH 1 OTA data (Figure S5) indicates that a new species forms with a time constant of $\sim$97 ps and decays with an accelerated lifetime of $\sim$2.9 ns. The latter is consistent with a previous report on the decay of the non-emissive excited-state of \rubpz{} in acidic solution ~\cite{Shinozaki1989}. The OTA spectra at 100 ps and 300 ps in Figure~\ref{fig:Figure 1}B show that this new species is spectrally similar to the $^{3}$MLCT$_{\text{bpz}}$ state, but the ESA at $\sim$350 nm blue-shifts and decreases in amplitude. This indicates an electronic reorganization at the ligand level, that we attribute to excited-state protonation. To directly capture this process from the perspective of the N site involved, we turn to element-specific N K-edge XAS.


\section{Site-specific electronic signatures of PCET}
\begin{figure}[H]
\centering
\includegraphics[width=1.0\textwidth]{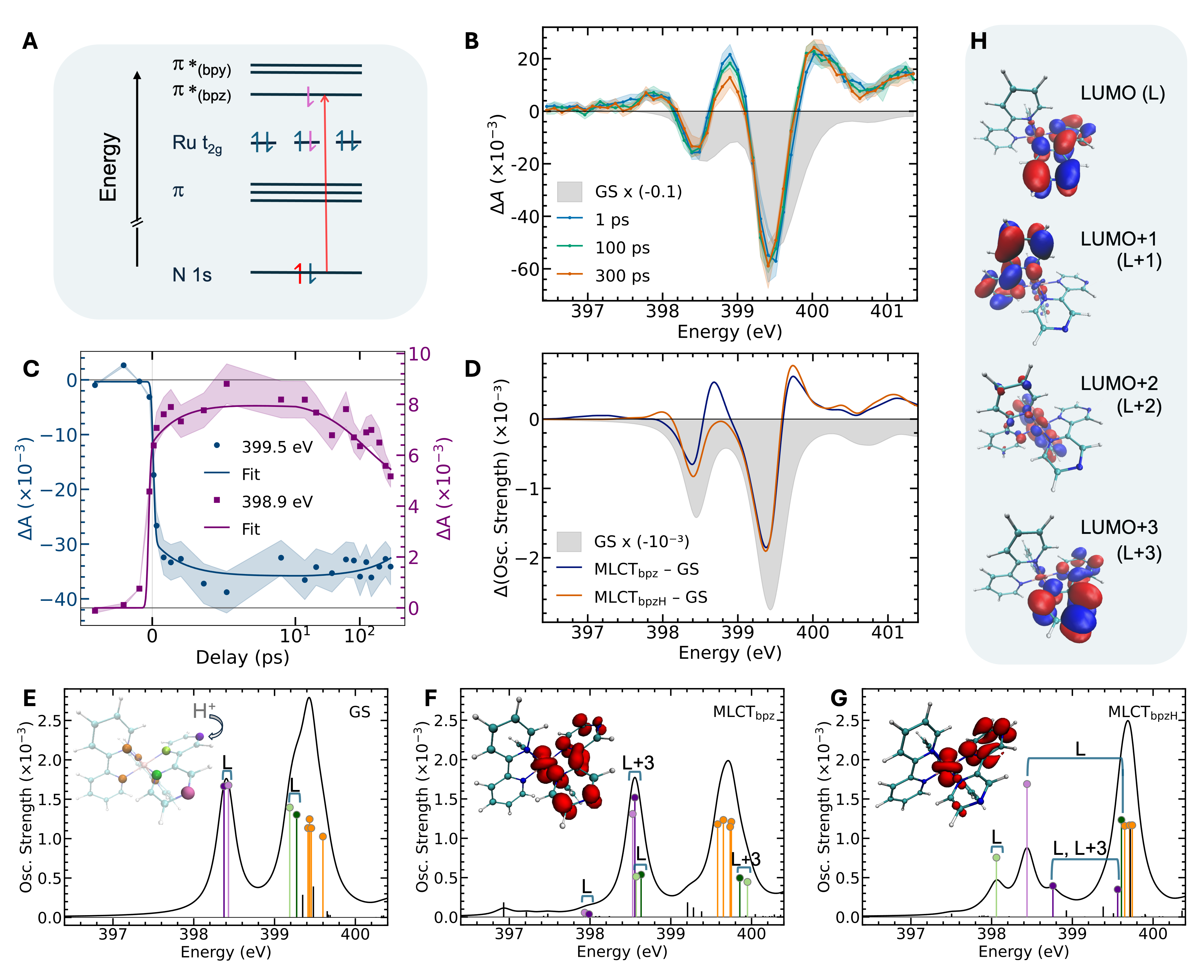}
\caption{Ultrafast N K-edge response of \rubpz{} and assignment with TDDFT. (A) Schematic molecular orbital diagram illustrating the N K-edge XAS process. The red vertical arrow shows a representative transition from the N 1s to a $\pi^{*}$ orbital  localized on the bpz ligand.  (B) Transient N K-edge spectra recorded at 1, 100, and 300 ps after 400 nm excitation at pH 1. The gray shaded area represent the negative scaled ground-state absorption. (C) Kinetic traces at 399.5 eV (blue, left axis) and 398.9 eV (magenta, right axis) with 95\% confidence intervals. Solid lines show the fits obtained from a sequential kinetic model, in which the time constants are fixed to the values obtained from the global analysis of the OTA data. For the GSB, the model includes a 2 ps component (VR) and a 3 ns component (decay of the excited-state). For the positive transient, an additional 97 ps component is included to account for excited-state protonation. 
(D) TDDFT calculations of the $^{1}$GS N K-edge spectra (gray fill, shown inverted) and transient spectra corresponding to the  $^{3}$MLCT${_\text{bpz}}$ (blue) and $^{3}$MLCT${_\text{bpzH}}$ (orange) states. 
 (E-G) Calculated N K-edge spectra (black, 0.25 eV Lorentzian-broadened) with underlying transitions for (E) $^{1}$GS, (F)  $\;^{3}\mathrm{MLCT}_{\mathrm{bpz}}$, and (G) $\;^{3}\mathrm{MLCT}_{\mathrm{bpzH}}$. Insets: (E) molecular structure indicating the color-coded individual Ns: bpy-N (orange), coordinated bpz-N (green), peripheral bpz-N (purple; dark purple marks the protonated N); (F,G) spin-density isosurfaces of the corresponding excited-states, plotted at an isovalue of $\pm$0.005 a.u. The color of the transition reflects the N site involved; labels “L” (LUMO), “L+1/2/3” denote the most contributing unoccupied orbitals. (H) Most relevant ground-state unoccupied molecular orbitals shown as isosurfaces with isovalue of $\pm$0.03 a.u., which are used to label the dominant features in panels E-G.}
\label{fig:Figure 2}
\end{figure}

The N K-edge XAS spectrum arises from N 1s core-level excitations to unoccupied  $\pi^*$ orbitals with N 2p character (Figure~\ref{fig:Figure 2}A). The steady-state N K-edge spectrum of \rubpz{}, plotted as inverted gray shading in Figure~\ref{fig:Figure 2}B), features two peaks at 398.5 and 399.5 eV. The low-energy band (398.5 eV) is assigned to transitions involving the 1s electrons of the peripheral bpz Ns, while the more intense, higher energy band (399.5 eV) is assigned to transitions involving the six Ns directly coordinated to Ru.~\cite{Jay2021,Ryland2024,AmyRuDPPZ}

Figure~\ref{fig:Figure 2}B also shows transient N K-edge spectra collected upon photoexcitation of \rubpz{} at pH 1.  The transient at 1 ps (blue) site-selectively probes the $^{3}$MLCT${_\text{bpz}}$ state and exhibits GSBs at 398.4 and 399.5 eV, with positive transients at 398.9 and 400.0 eV. As supported by further analysis (Figure S8 and Table S1), these features arise from a blue-shift and reduced intensity of the high-energy band, as well as a slight blue-shift of the low-energy band. A weaker positive transient at 397.9 eV is also observed.

Transient spectra at 100 ps (green) and 300 ps (orange) show a subsequent decrease in the positive transient at 398.9 eV, corresponding to the blue-shifted low-energy band assigned to transitions of peripheral bpz Ns. We attribute this decrease to protonation of a peripheral bpz N. Kinetic data (Figure~\ref{fig:Figure 2}C) are consistent with the reshaping of the 300~ps signal with respect to the 1~ps signal, indicating formation of a new species. The GSB measured at 399.5 eV (assigned to transitions of the Ru-coordinated Ns insensitive to peripheral protonation) appears instantaneously upon photoexcitation and remains constant. In contrast, the positive signal at 398.9 eV decays over time. These dynamics are consistent with the kinetic model derived from the OTA data (solid lines) and are absent at pH 7, where excited-state protonation is not expected (Figure S7).



To interpret the N K-edge XAS data, we computed N 1s X-ray absorption spectra for the ground-state ($^{1}$GS), the lowest triplet MLCT state ($^{3}$MLCT${_\text{bpz}}$), and the protonated triplet MLCT state ($^{3}$MLCT${_\text{bpzH}}$) using time-dependent density functional theory (TDDFT). As detailed in the SI Section 6.3, each spectrum was averaged over 20 QM/MM snapshots to explicitly account for solute–solvent interactions.  The simulated ground-state and transient spectra (Figure~\ref{fig:Figure 2}D) reproduce the key experimental features, supporting the assignment of the 1–300 ps changes in our XAS data to excited-state protonation. TDDFT calculations of valence excitations on the same snapshots reproduce the transient changes assigned to protonation in the OTA measurements (Figure S6). While the valence TDDFT spectra are difficult to interpret because of the high degree of delocalization and mixed character of the excitations, N K-edge spectroscopy provides a clearer, orbital-level picture, which we describe in the following.

Figures~\ref{fig:Figure 2}E-G display the calculated transitions for a representative snapshot of each electronic state. Transitions are color-coded by N type (green for coordinated bpz Ns, purple for peripheral bpz Ns, and orange for bpy Ns) as shown in the inset of ~\ref{fig:Figure 2}E. Transitions are labeled by the target virtual molecular orbitals shown in ~\ref{fig:Figure 2}H. We provide detailed assignments of individual N transitions and a full set of orbitals and their compositions in SI Section 7.

In the $^{1}$GS (Figure~\ref{fig:Figure 2}E), the low-energy band arises from N\,1s\,$\rightarrow$\,LUMO and LUMO+3 transitions of peripheral bpz Ns, with both orbitals localized on the bpz ligand. The higher energy band comprises transitions from the 1s orbitals of (i) the coordinated bpz Ns to the LUMO and LUMO+3  and (ii) bpy Ns to the bpy-localized LUMO+1 and LUMO+2. Because coordination to Ru stabilizes the N 1s levels, transitions from Ru-coordinated bpz N atoms into the same LUMO and LUMO+3 appear at higher energy than those from peripheral bpz N atoms.

In the $^{3}$MLCT${_\text{bpz/bpzH}}$ states, an electron is promoted from the ground-state HOMO to the ground-state LUMO. In the open-shell treatment this leads to substantial redistribution of valence spin–orbital energies, yet the orbitals remain generally localized on either bpy or bpz. Transitions from bpy N 1s orbitals occur predominantly into bpy-localized  $\pi^*$ orbitals (nominally corresponding to the ground-state LUMO+1 and LUMO+2), whereas transitions from bpz N 1s orbitals occur predominantly into bpz-localized  $\pi^*$ orbitals (nominally LUMO and LUMO+3).


We first examine the excited-state changes of the high-energy XAS peak, which is dominated by the bpy Ns transitions, color-coded in orange. In the $^{3}$MLCT${_\text{bpz}}$ state (Figure ~\ref{fig:Figure 2}F), the bpy N\,1s orbitals are stabilized by approximately 0.7 eV due to enhanced $\sigma$-donation to the formally Ru$^{\text{III}}$. The bpy $\pi^*$ orbitals are also stabilized, though less strongly. Accounting for Coulomb and exchange interactions, the net effect is a predicted $\sim$0.2 eV blue-shift of the high-energy XAS peak, consistent with our experiment, as well as with prior measurements on photoexcited \rubpy~\cite{Jay2021}. In the $^{3}$MLCT${_\text{bpzH}}$ state (Figure~\ref{fig:Figure 2}G), the increased positive charge of the complex induces an overall stabilization of all molecular orbitals. Unsurprisingly, due to the distal proximity of the bpy ligands to the protonation site, there is no net change on the bpy N 1s $\rightarrow$ $\pi^*$ transitions between the protonated and unprotonated $^{3}$MLCT states. Thus, the higher energy XAS peak is mostly unaffected by protonation. 


We next analyze the excited-state evolution of the low-energy XAS peak, which arises from transitions of the bpz N atoms. In the $^{3}$MLCT$_\text{bpz}$ state (Figure~\ref{fig:Figure 2}F), the N 1s orbitals of both Ru-coordinated  and peripheral  bpz Ns are destabilized by ~0.5 eV due to increased electron density on the ligand. For the coordinated Ns (green), this destabilization is the main factor determining the red-shift of their transitions. For the peripheral Ns (purple), however, the dominant transitions target bpz $\pi^*$ orbitals corresponding to the ground-state LUMO+3, whereas the LUMO is the dominant contributor in the ground-state. This substitution of the target orbital from LUMO to higher-lying LUMO+3 offsets the N 1s destabilization, yielding a small net blue-shift for these transitions. Overall, TDDFT predicts a $\sim$0.1 eV blue-shift of the low-energy XAS peak, in agreement with experiment.

In the $^{3}$MLCT$_\text{bpzH}$ state (Figure~\ref{fig:Figure 2}G), electron density localizes on the pyrazine ring containing the protonated N. Consequently, N atoms on the unprotonated pyrazine ring (dark green and light purple) exhibit transitions similar in energy and intensity to those in the ground-state. The protonated N (dark purple) shows blue-shifted transitions due to 1s orbital stabilization by the added positive charge, along with reduced oscillator strengths relative to the $^{1}$GS and $^{3}$MLCT$_\text{bpz}$ states. Collectively, these effects reduce the intensity of the low-energy XAS peak without generating a positive transient on its high-energy shoulder, whereas in the $^{3}$MLCT\textsubscript{bpz} state, a positive transient appears due to the peak's blue-shift. These trends are consistent with the experimentally observed reduction of the 398.9 eV transient upon protonation (Figure~\ref{fig:Figure 2}C).


An additional calculated low-energy transition involving the coordinated bpz N of the protonated ring (light green) is consistent with the positive transient at 397.9 eV. The calculations do not reproduce a corresponding transient in the $^{3}$MLCT$_\text{bpz}$ state, possibly because they underestimate the intensity of transitions to the LUMO in this region. A closer comparison of the measured signals at 1 ps ($^{3}$MLCT$_\text{bpz}$) and 300 ps ($^{3}$MLCT$_\text{bpzH}$) reveals a reshaping of the 397.9 eV feature, supporting its assignment to distinct electronic origins in the two states. Finally, weak transitions predicted below 398.4 eV correspond to N 1s → Ru 4d $(\pi)$ excitations that borrow dipole intensity via Ru d–N p orbital mixing, but these are not observed experimentally. 

Overall, N K-edge XAS, supported by TDDFT, allows us to track both electron localization and  site-specific protonation in the PCET reaction of \rubpz. 

\section{Solvent Reorganization coupled to PCET}

\begin{figure}[H]
\centering
\includegraphics[width=1\linewidth]{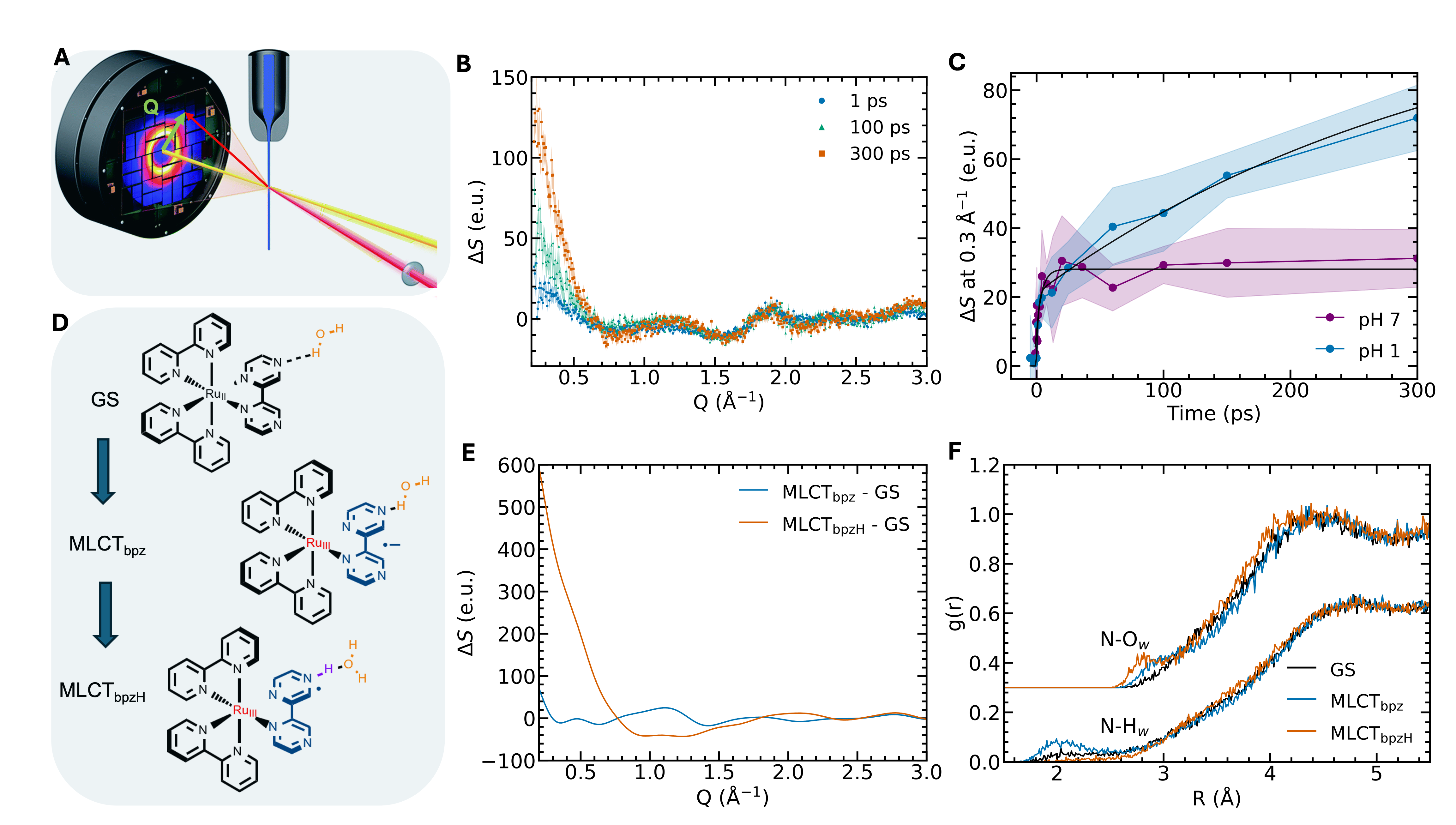}
\caption{%
Ultrafast X-ray scattering reporting on solvent reorganization upon photoexcitation of \rubpz{}.    
(A) Schematic illustration of the experimental setup. Following 530\,nm excitation, elastic X-ray scattering was recorded on a forward detector and radially integrated as a function of the scattering vector Q.  (B) Experimental difference scattering signals ($\Delta S$) at 1, 100, and 300\,ps, showing a growing low-Q feature on the timescale of protonation.  (C) Kinetic traces of the low-Q scattering signal (integrated over 0.25–0.35\,Å$^{-1}$) at pH 1 (blue) and pH 7 (magenta) with 95\% confidence intervals. Fits (black solid lines) use single- or double-exponential rise models and are a guide to the eye.  
(D) Structural evolution pathway of the local hydrogen bonding network for photoexcited \rubpz{}.  (E) Simulated difference scattering curves for the $^{3}$MLCT${_\text{bpz}}$ (blue) and $^{3}$MLCT${_\text{bpzH}}$ (orange), reproducing the low-Q increase upon protonation.  (F) Radial distribution functions (RDFs) between N atoms and water hydrogens (N–H$_\mathrm{w}$) and oxygens (N–O$_\mathrm{w}$), calculated for the GS, MLCT${_\text{bpz}}$, and MLCT${_\text{bpzH}}$ states. The N–O$_\mathrm{w}$ curves are vertically offset by 0.3 for clarity.
}
\label{main_xss}
\end{figure}

To shed light on the role of the aqueous environment in the PCET reaction, we performed a separate time-resolved XSS experiment (Figure~\ref{main_xss}A).  Figure~\ref{main_xss}B shows transient scattering curves measured at 1~ps, 100~ps, and 300~ps after photoexcitation of \rubpz{} at pH 1. As established in previous studies~\cite{Biasin2016, IheeReview}, transient XSS signals arise from photoinduced changes in intra-molecular (solute–-solute) and inter-molecular (solute–-solvent) atomic pair distances, as well as structural changes in the bulk solvent due to transient heating from excited-state relaxation. The latter contribution was subtracted using standard procedures~\cite{KasperHeat}. As is typical for Ru-polypyridyl complexes and confirmed by TDDFT (SI Section 8.1), the intra-molecular bond distances of \rubpz{} remain nearly unchanged between the ground- and lowest $^{3}$MLCT$_\text{bpz/bpzH}$ states. Thus, the observed transient scattering signal mainly reflects changes in solute–-solvent distances, indicating reorganization of the solvation shell.

At 1~ps, the signal captures the initial solvent reorganization upon population of the $^{3}$MLCT$_\text{bpz}$ state. The positive transient at low momentum transfer (Q) suggests a shortening of solute–-solvent distances, likely reflecting strengthened hydrogen bonding between the solvent and the reduced bpz ligand, as observed previously~\cite{BiasinNatChem}. Notably, the low-Q transient signal continues to grow over 1–300 ps,  consistent with the timescale of excited-state protonation. This continued growth at pH 1 is further shown in the kinetic trace in Figure~\ref{main_xss}(C). In contrast, at pH 7, the low-Q signal rapidly plateaus after an initial rise. These observations support an assignment of the evolving low-Q signal to structural changes driven by excited-state protonation, and are consistent with increased local electron density relative to the ground and unprotonated states. 

To further interpret the measured difference scattering signals, we performed classical molecular dynamics (MD) simulations, which provide the statistical sampling necessary for an accurate description of solvation dynamics. Solute--solvent radial distribution functions (RDFs) and corresponding difference scattering signals were calculated for the ground-state $^{1}$GS, $^{3}$MLCT${_\text{bpz}}$, and $^{3}$MLCT${_\text{bpzH}}$ state, following an established procedure~\cite{Dohn_2015}. As shown in Figure~\ref{main_xss}E, the calculated transient scattering signals reproduce the increase in the low-Q positive feature upon protonation. 

Having verified that the simulations reproduce our experimental observables, we next examine the RDFs to gain atomistic insight into solvation-shell changes across the three electronic states. Figure~\ref{main_xss}F shows the N-H${_\text{w}}$ and N-O${_\text{w}}$ RDFs, where N comprises all N atoms of \rubpz{}, while  H${_\text{w}}$ and O${_\text{w}}$ represent the H and O atoms in water, respectively.

In the $^{3}$MLCT${_\text{bpz}}$ state, both N–H${_\text{w}}$ (2.0~Å) and N–O${_\text{w}}$ (2.8~Å) first-shell peaks increase relative to the ground state, consistent with strengthened hydrogen bonding to the reduced bpz ligand~\cite{Drummer2022,Wenger}. Upon protonation, the N–H${_\text{w}}$ peak decreases while the N–O${_\text{w}}$ peak increases, reflecting H-bond formation between the protonated bpz N and nearby O${_\text{w}}$. This transition from N$\cdots$H${_\text{w}}$O${_\text{w}}$ to NH$\cdots$O${_\text{w}}$ interactions (Figure~\ref{main_xss}D) implies a reorientation of water molecules around the protonated pyrazine. The rearrangement increases local electron density in the first solvation shell, contributing to the enhanced low-Q scattering signal.

Experimentally, solvation of individual atoms cannot be resolved. Analysis of MD simulations (SI Section 8.3) suggests that, while water also approaches the bpy ligands due to their increased positive charge in both MLCT${_\text{bpz/bpzH}}$ states~\cite{Maroncelli}, more than half of the low-Q signal arises from reorganization around the bpz ligand. In the MLCT${_\text{bpz}}$ state, reorientation of water near bpz C - qualitatively consistent with prior reports~\cite{Mai_2025} - compensates for the strengthening of H-bonding to the bpz Ns, yielding a negligible net solute-–solvent signal. This is in contrast with the small positive signal observed experimentally, and this discrepancy will be addressed in future work. In the MLCT${_\text{bpzH}}$ state, both C and N of the bpz ligand gain electron density through the approach of solvent Os, with most reorganization localized around the protonated pyrazine ring.

\section{Conclusions and Implications}
To capture the coupled motions of electrons, protons, and solvent in PCET at the atomic scale, we combined femtosecond optical spectroscopy, site-specific N K-edge X-ray absorption, and time-resolved X-ray scattering. 

In \rubpz{}, transient N K-edge features, interpreted with TDDFT, directly reveal the population of the lowest $^{3}$MLCT${_\text{bpz}}$ state and the subsequent protonation of a peripheral bpz N within $\sim$100 ps. These measurements provide a site-specific, orbital-level picture of protonation and its impact on local electronic structure, distinct from charge-transfer effects. This framework opens pathways to resolve long-standing mechanistic questions, including whether PCET proceeds sequentially or concertedly, and whether proton and electron motions are synchronous or asynchronous.

Time-resolved XSS, supported by MD simulations, directly tracks the shortening of solute–solvent interatomic distances, mainly driven by the reorganization of solute--solvent H-bonding. Upon excitation, the H-bonding between the water molecules and the peripheral Ns of the reduced bpz ligand is strengthened, while protonation triggers the transition from N$\cdots$HO to NH$\cdots$O interactions at the protonated N. Future ultrafast studies may capture how solvent reorganization facilitates — or even gates — the protonation event. 

Together, these measurements provide the first integrated view of electronic and solvent dynamics during a PCET reaction. Our multimodal framework establishes a path to disentangle PCET mechanisms, informing the rational design of PCET in catalysis, artificial photosynthesis, and biological energy conversion.

\section{Acknowledgments}
We thank Prof. Munira Khalil for support during the chemRIXS experiment and Prof. Kelly J. Gaffney for useful discussions. This work was supported by the U.S. Department of Energy, Office of Science, Basic Energy Sciences, Chemical Sciences, Geosciences, and Biosciences Division, Condensed Phase and Interfacial Molecular Science (CPIMS), FWP 16248 (E.~B. and A.~K.), and Atomic, Molecular, and Optical Sciences (AMOS) Programs, FWP 72684 (S.~G. and N.~G.), at Pacific Northwest National Laboratory (PNNL). B.I.P., E.S.R., N.P.R., and A.A.C. were supported by the U.S. Department of Energy (DOE), Office of Basic Energy Sciences, Division of Chemical Sciences, Geosciences and Biosciences, through SLAC National Accelerator Laboratory under Contract No. DE-AC02-76SF00515. C.~B.~L. gratefully acknowledges a SNSF Ambizione Grant (Grant Number 193436). Use of the Linac Coherent Light Source (LCLS), SLAC National Accelerator Laboratory, is supported by the U.S. Department of Energy, Office of Science, Office of Basic Energy Sciences under Contract No. DE-AC02-76SF00515. A portion of this research was performed on project awards (60604, 61181) at the Environmental Molecular Sciences Laboratory (EMSL), a DOE Office of Science User Facility sponsored by the Biological and Environmental Research program located at Pacific Northwest National Laboratory (PNNL). This research also gratefully acknowledges computational resources provided by the PNNL Institutional Computing (PIC) Program and National Energy Research Scientific Computing Center (NERSC). PNNL is operated by Battelle Memorial Institute for the United States Department of Energy under DOE Contract No. DE-AC05-76RL1830. NERSC is a U.S. Department of Energy Office of Science User Facility operated under Contract No. DE-AC02-05CH11231.

\bibliographystyle{unsrt}
\bibliography{bibliography}

\end{document}

%% file: authors.tex
\author[1,2]{Abdullah Kahraman}
\author[2]{Michael Sachs} 
\author[1,9]{Soumen Ghosh}
\author[3]{Benjamin I. Poulter}
\author[4]{Estefanía Sucre-Rosales}
\author[2]{Elizabeth S. Ryland}
\author[2]{Douglas Garratt}
\author[2]{Sumana L. Raj} 
\author[2]{Natalia Powers-Riggs}
\author[4]{Subhradip Kundu}
\author[5]{Christina Y. Hampton}
\author[5]{David J. Hoffman}
\author[5]{Giacomo Coslovich} %
\author[5]{Georgi L. Dakovski} %
\author[5]{Patrick L. Kramer}
\author[5]{Matthieu Chollet} %
\author[5]{Roberto A. Mori} %
\author[5]{Tim B. van Driel} %
\author[6]{Sang-Jun Lee}
\author[5]{Kristjan Kunnus}
\author[2]{Amy A. Cordones}
\author[2,5]{Robert W. Schoenlein}
\author[4]{ Eric Vauthey}
\author[7]{Amity Andersen}
\author[1,3]{Niranjan Govind}
\author[8*]{Christopher Larsen}
\author[1,2*]{Elisa Biasin}

\affil[1]{Physical Sciences Division, Physical and Computational Sciences Directorate, Pacific Northwest National Laboratory, Richland, WA 99354, USA}
\affil[2]{Stanford PULSE Institute, SLAC National Accelerator Laboratory, Menlo Park, CA 94025, USA}
\affil[3]{Department of Chemistry, University of Washington, Seattle, Washington 98195, USA}
\affil[4]{Department of Physical Chemistry, University of Geneva, CH-1211 Geneva, Switzerland}
\affil[5]{Linac Coherent Light Source, SLAC National Accelerator Laboratory, Menlo Park, CA 94025, USA.}
\affil[6]{Stanford Synchrotron Radiation Lightsource, SLAC National Accelerator Laboratory, Menlo Park, CA 94025, USA.}
\affil[7]{Environmental Molecular Sciences Laboratory, Earth and Biological Sciences Directorate, Pacific Northwest National Laboratory, Richland, WA 99354, USA }

\affil[8]{School of Chemical Sciences, University of Auckland, Auckland 1010, New Zealand}
\affil[9]{Present Address: Department of Chemistry, Indian Institute of Technology Madras, Chennai, 600036 India}
